# The effect of remote work on urban transportation emissions: evidence from 141 cities

**Sophia Shen**
Department of Urban Studies and Planning
Massachusetts Institute of Technology, Cambridge, MA 02139, USA
Email: s_shen@mit.edu

**Xinyi Wang**
Department of Urban Studies and Planning
Massachusetts Institute of Technology, Cambridge, MA 02139, USA
Email: xinyi174@mit.edu

**Nicholas Caros**
Department of Civil and Environmental Engineering
Massachusetts Institute of Technology, Cambridge, MA 02139, USA
Email: caros@mit.edu

**Jinhua Zhao**
Department of Urban Studies and Planning
Massachusetts Institute of Technology, Cambridge, MA 02139, USA
Email: jinhua@mit.edu

**ABSTRACT**

The overall impact of working from home (WFH) on transportation emissions remains a complex issue, with significant implications for policymaking. This study matches socioeconomic information from American Community Survey (ACS) to the global carbon emissions dataset for selected Metropolitan Statistical Areas (MSAs) in the US. We analyze the impact of WFH on transportation emissions before and during the COVID-19 pandemic. Employing cross-sectional multiple regression models and Blinder-Oaxaca decomposition, we examine how WFH, commuting mode, and car ownership influence transportation emissions across 141 MSAs in the United States. We find that the prevalence of WFH in 2021 is associated with lower transportation emissions, whereas WFH in 2019 did not significantly impact transportation emissions. After controlling for public transportation usage and car ownership, we find that a 1% increase in WFH corresponds to a 0.17 kilogram or 1.8% reduction of daily average transportation emissions per capita. The Blinder-Oaxaca decomposition shows that WFH is the main driver in reducing transportation emissions per capita during the pandemic. Our results show that the reductive influence of public transportation on transportation emissions has declined, while the impact of car ownership on increasing transportation emissions has risen. Collectively, these results indicate a multifaceted impact of WFH on transportation emissions. This study underscores the need for a nuanced, data-driven approach in crafting WFH policies to mitigate transportation emissions effectively.

**INTRODUCTION**

There has been substantial debate around the impact of working from home (WFH) on transportation carbon emissions relative to employer premises-based work (*1*). WFH eliminates the need to travel to the office, thus reducing the transportation-related emissions from commuting. Yet, people who work from home still engage in leisure and household activities, and the evidence of overall reductions in personal travel due to WFH is mixed and highly variable (*2-3*). Moreover, WFH may induce secondary changes to transportation behavior, such as changes in mode choice, that contribute to increased carbon emissions. Sepanta and O'Brien (*4*) provide an in-depth exploration of the different mechanisms by which WFH affects carbon emissions. As a result, the impact of WFH on transportation emissions remains relatively uncertain despite the important policy implications.

Recently, a new data source has emerged that may shed light on this issue. Huo et al. (*5*) publish detailed carbon emissions data, broken out by sector, for 1500 cities across the world. The emissions sectors include ground transportation, residential buildings, power generation, aviation, and industry processes. The emissions data spans from the start of 2019 through the end of 2021, a period during which travel behavior varied significantly due to the COVID-19 pandemic. The spatial extent of the emissions data makes it possible to disentangle the confounding effects of WFH on transportation emissions. One might expect that WFH, by reducing commuting demand, should produce fewer transportation emissions. At the same time, the communities where WFH is prevalent could also be places with a host of socioeconomic characteristics that may influence transportation emissions. Disentangling the confounding effects presents an opportunity to gain insights for important policy decisions.

The American Community Survey (ACS) (6-7), conducted annually by the U.S. Census Bureau, provides detailed data on demographics, housing characteristics, commute behavior, and employment statistics for communities across the United States. Given the abundance of socioeconomic variables available from ACS, we begin with a conceptual framework to focus on the impact of transportation emissions from WFH. We merge the ACS data with the emissions data to analyze the complex impact of WFH on transportation emissions.

One unique feature of studying the impact of WFH on transportation emissions is that reverse causality is unlikely to occur, as residents would not be expected to change their WFH behavior in response to changes in transportation emissions. In addition, the unique opportunity for analyzing the impact of WFH on transportation emissions from 2019 to 2021 is that WFH increased substantially following the onset of the pandemic; WFH participation in the US went from less than 5% of worked hours in 2018 to more than 30% of worked hours in 2021 (*8*). The pandemic's impact on transportation demand is equally dramatic: public transportation usage declined by 40% from 2019 to 2021 in the United States. The magnitude of the change and impact gives us an unprecedented opportunity to disentangle the multi-faced relationships and gain policy insights.

We use a cross-sectional Ordinary Least Squares (OLS) regression approach to analyze the impact of WFH on transportation emissions across 141 MSAs in the US. We analyze the impact in relation to other travel behavior choices, such as commute mode and car ownership, that could also affect transportation emissions. The unique contribution of this paper is to synthesize carbon emissions data for over 100 cities with aggregated sociodemographic and travel behavior data to explore the impact of WFH on transportation emissions. By using the OLS approach with Blinder-Oaxaca decomposition, we disentangle the confounding effects of WFH, commuting modes, and car ownership on transportation emissions.

Our findings indicate that, while WFH did not have a significant impact on transportation emissions before the pandemic, the prevalence of WFH had a measurable impact on reducing transportation emissions in 2021. Specifically, we find that a 1% increase in WFH corresponds to a 0.17 kilogram or 1.8% reduction of daily average transportation emissions per capita, while controlling for the effects of public transportation and car ownership. From the Blinder-Oaxaca decomposition, we show that WFH is the main driver in reducing transportation emissions per capita during the pandemic. Our results indicate that the downward influence of public transportation on transportation emissions has declined in 2021. Simultaneously, the impact of car ownership on increasing transportation emissions rose.

**LITERATURE REVIEW**

WFH has been extensively studied, especially its impact on travel behavior, commute pattern, energy usage, and carbon emissions. Relevant literature is summarized in the subsections below.

**WFH and Travel Behavior**

In the early stages of the information age, Hong (*2*) explored remote work's impact on travel behavior. Despite early optimism that telecommunications would reduce physical travel, empirical evidence revealed a significant increase in passenger travel. However, WFH saw widespread adoption during the COVID-19 pandemic. Zheng et al. (*9*) studied the pandemic's impact on travel behavior in Massachusetts, particularly noting a shift towards increased car commuting as pre-pandemic routines resumed post-vaccine availability in fall 2021.

Migration and commute patterns were significantly impacted by the pandemic. Ramani and Bloom (*10*) observe a shift in household and business relocation from urban cores to suburban and exurban areas in major U.S. cities, reducing the need for daily work commutes and making longer commutes more feasible. Similarly, Asmussen et al. (*11*) found that 20% of telework decisions led to residential moves impacting commute distances, while 80% of new teleworkers adjusted their telework habits based on pre-existing residential locations, with varying degrees of telework affecting commute vehicle miles traveled. Reiffer et al. (*12*) noted a marked increase in telecommuting during the pandemic, especially among new teleworkers influenced by household dynamics, like childcare.

Haleform et al. (*13*) systematically reviewed literature highlighting WFH's environmental, social, and economic benefits. They also noted challenges posed by urban patterns and lifestyle adjustments, particularly highlighting the need for frequent WFH participation to achieve significant reductions in travel distance. Magassy et al. (*14*) predicted a sustained 30% decrease in U.S. transit ridership post-pandemic. They find that minority groups and residents in higher-density areas are more likely to return to public transit use post-pandemic.

**WFH and Emissions**

The growing use of information and communication technologies reshaped daily work patterns and influenced emissions. Cerqueira and Motte-Baumvol (*15*) explored how this shift impacted travel behavior, workplace diversification, and environmental outcomes from 2002 to 2017. They observed that hybrid and remote workers reported higher emission levels compared to those with a single workplace.

O'Brien and Aliabadi (*1*) reviewed quantitative studies on the impact of telecommuting on energy consumption. Despite remote work being touted as a sustainable alternative to traditional office-based work, findings indicated a mix of potential energy and GHG emission reductions, alongside increases due to rebound effects like changes in home energy use, transportation decisions, and consumer behavior. Similarly, Sepanta and O'Brien (*4*) looked at remote work's impact on energy use across offices, homes, and transportation. Despite its widespread adoption, remote work's impact on energy consumption and carbon emissions remains ambiguous due to the complexity of remote work's domains, with empirical evidence showing mixed outcomes and potential rebound effects that can negate initial energy savings.

Wu et al. (*16*) found that, during the pandemic, widespread WFH reduced GHG emissions by 29%. This is primarily by reducing commuting and workplace emissions, despite increased residential emissions, underscoring WFH or hybrid arrangements as effective GHG reduction strategies. However, post-pandemic, WFH's environmental impact on emissions is nuanced. Caros et. al (*17*) and Caros (*18*) uses national survey data and mobile trace data to estimate Chicago's commuting patterns, revealing that remote workers in the US spend about one-third of their WFH hours outside the home at a "third place", such as cafés and co-working spaces, offsetting expected reductions in congestion and emissions. They show that ignoring these "third places" underestimates commute-based carbon emissions by 24%.

In summary, previous literature reveals a complex, evolving, and multifaceted set of factors contributing to the relationship between WFH and travel behavior, commute pattern, energy usage

*Shen, Wang, Caros, and Zhao*

and carbon emissions. The literature also underscores the need for additional research to gain a deeper understanding in this important area.

**METHODS**

**Conceptual Framework**

While many factors can impact transportation emissions, the focus of this study is to quantify the impact of WFH on transportation emissions. Structurally, several factors might produce cross-sectional differences in transportation emissions. These factors include population size and density, transportation infrastructure, and other socioeconomic characteristics, such as income and housing affordability. We are less focused on explaining the cross-sectional differences of transportation emissions in totality as there is a deep understanding of the relationship between transportation emissions and different structural factors in the literature (*19*). The limitation of the data coverage of 141 cities also constrains the number of independent variables to maintain statistical power. Our focus on the impact of the pandemic leads us to explore cross-sectional drivers that have shifted significantly over the course of the pandemic that would contribute to a relative change in carbon emissions from transportation.

The dramatic increase in WFH induced changes related to transportation behaviors and choices, including a reduction in commuting to the workplace, an increase in transportation demand to "third places," decreased demand for public transportation, and increased demand for driving alone (*11*, *17*). The effects are likely to persist, at least into the short and medium-term future, as WFH levels have remained elevated well after the public health crisis has subsided (*8*). In addition, WFH patterns are heterogeneous across MSAs in the US (*8*). This study, therefore, investigates transportation emissions as a function of WFH participation, commuting mode choice, and vehicle ownership. Employing cross-sectional regression models across 141 MSAs in the United States, this analysis aims to elucidate the relationship between these factors.

**Data Collection**

*Carbon Monitor Cities*

For transportation emissions data, we use the Carbon Monitor Cities (CMC) dataset published in Scientific Data (5). The CMC provides near-real-time daily estimates of $CO_2$ emissions from over 1,500 cities worldwide, spanning from January 2019 to December 2021. The emissions data is organized into five distinct sectors: transportation, residential and commercial buildings, industrial processes, power, and aviation. The comprehensive coverage of the CMC dataset, coupled with its detailed sectoral breakdown and data continuity throughout the pandemic, allows us to gain insights into urban carbon footprints.

Daily emissions for ground transportation are estimated using the TomTom Congestion Index and the EDGAR on-road emissions dataset. Traffic volume data is used to allocate the EDGAR on-road emissions to each day to obtain estimates of daily ground transportation emissions. To ensure robustness in our analysis and mitigate seasonal fluctuations, we aggregate the daily transportation emissions for the relevant geographic areas over the years of 2019 and 2021 to obtain the total annual transportation emissions for both years.

*American Community Survey*

For data on transportation demand and socioeconomic characteristics, we use the American Community Survey (ACS), conducted annually by the U.S. Census Bureau (6-7). We focus on three categories of ACS respondent information: demographics, housing characteristics, and employment statistics. We extracted population, household income, employment industry, commute mode to work, car ownership, and WFH from ACS (**Table 1**). The ACS data is collected on an ongoing basis and released in both one-year and five-year estimates. The 1-year estimates provide frequent updates for metropolitan areas with 65,000 or more residents, while 5-year estimates increase statistical reliability by combining 5 years of collected data. We use the one-year estimate ACS data in 2019 and 2021 for our analysis. ACS data for 2020 was unavailable due to significant reduction with response rate during the pandemic.

*Merging Datasets*

As we aim to establish the relationship between transportation emissions and a host of demographic, socioeconomic and transportation demand characteristics, we arrive at the issue of defining a "city" or an "urban area." The CMC dataset uses Functional Urban Areas (FUA) and Global Administrative Areas (GADM), not US counties or MSAs, as the definition of a city. FUAs are designed to capture the economic and social functional reach of cities beyond their administrative boundaries, while GADM provides a more traditional, politically-defined boundary. These two classifications are applied globally, allowing for a consistent approach to compare urban emissions across different countries and regions.

The ACS dataset classifies its data by Metropolitan Statistical Area (MSA). MSAs are used in the United States to delineate regions that are socially and economically integrated with a core urban area, typically defined by population density and commuting patterns. An MSA encompasses not just a central city but also its surrounding suburbs and exurbs that have a high degree of interaction with the urban core.

In order to combine the transportation emission data from CMC and the socioeconomic data from ACS, we need to provide a mapping of FUA and GADM areas to the MSAs. The Organization for Economic Co-operation and Development (OECD) provides a crosswalk between FUAs and the corresponding countries in the US (*20*), allowing us to map FUAs onto US counties. Each MSA encompasses multiple counties as detailed in the crosswalk between MSAs and counties (*21*) provided by the United States Bureau of Labor Statistics. Given the linkages to US counties for both MSAs and FUAs, we use counties as the common unit of analysis to map FUAs to MSAs.

Emissions data from Carbon Monitor Cities is available for 211 FUAs in the US. While FUA and MSA are similar in many cases in the US, there could still be significant differences between the definitions when we compare the county compositions for FUA and MSA. Therefore, for each FUA, we examine the counties associated with that FUA and compare the counties associated with the corresponding MSA. If more than 33% of the counties in either the FUA or MSA are not included in the corresponding MSA or FUA, then the pair is excluded. This threshold ensures that the geographical scope of the FUAs and MSAs is sufficiently aligned for accurate mapping. By requiring that the geographies of the FUAs and MSAs do not deviate significantly from one another, we ensure that the socioeconomic data and emissions profiles are representative of a consistent group. After this mapping and cleaning process, we retain 141 FUAs, out of the original 211 FUAs, and their corresponding 141 MSAs. Our retained areas cover 192 million people in 2021, which is approximately 58% of the total US population. Out of the 141 retained FUAs, 112 of them (79%) have a perfect match with the corresponding MSA.

**Dependent Variable**

We designate the daily average of annual transportation emissions per capita, from CMC, as our dependent variable. We calculate this for 2019 and 2021 to analyze the impact of the pandemic. We use transportation emissions per capita because our empirical analysis reveals a clear association between daily average transportation emissions and the population size across MSAs. This is sensible as larger population size is associated with more transportation activity. To mitigate the inherent population size bias in daily average transportation emissions, we adjust the dependent variable to be the daily average transportation emissions per capita by dividing the daily average transportation emissions by the population of the corresponding MSA in the given year.

**Explanatory Variables**

*Work From Home (WFH)*

We measure WFH using the percentage of workers who participate in WFH, extracted from the ACS one-year estimates in 2019 and 2021. This percentage is part of the overall commute mode shares, which include modes such as public transportation and driving alone. The cross-MSA distribution of WFH for 2021 can be seen in **Figure 1**.

*Commute Mode*

We measure commute mode using the percentage of workers commuting to work through public transportation, extracted from the ACS one-year estimates in 2019 and 2021. The commute modes from ACS are categorized into six categories: driving alone, carpooling, public transportation, walking, other means, and working from home. Public transportation includes commuting to work via subway, elevated rail, long-distance train, commuter rail, light rail, streetcar, trolley, or ferryboat. Driving alone refers to commuting in a car, truck, or van with only one occupant, and the data does not distinguish between electric and gasoline-powered vehicles. Carpooling is defined as commuting in a car, truck, or van with two or more occupants. Since working from home is included as a commute mode, the definition of other commute modes, such as driving alone and carpooling, differs from the traditional commute mode share definition. To address this issue, we created new variables that rebase the share of each traditional mode out of the total traditional modes, excluding % WFH.

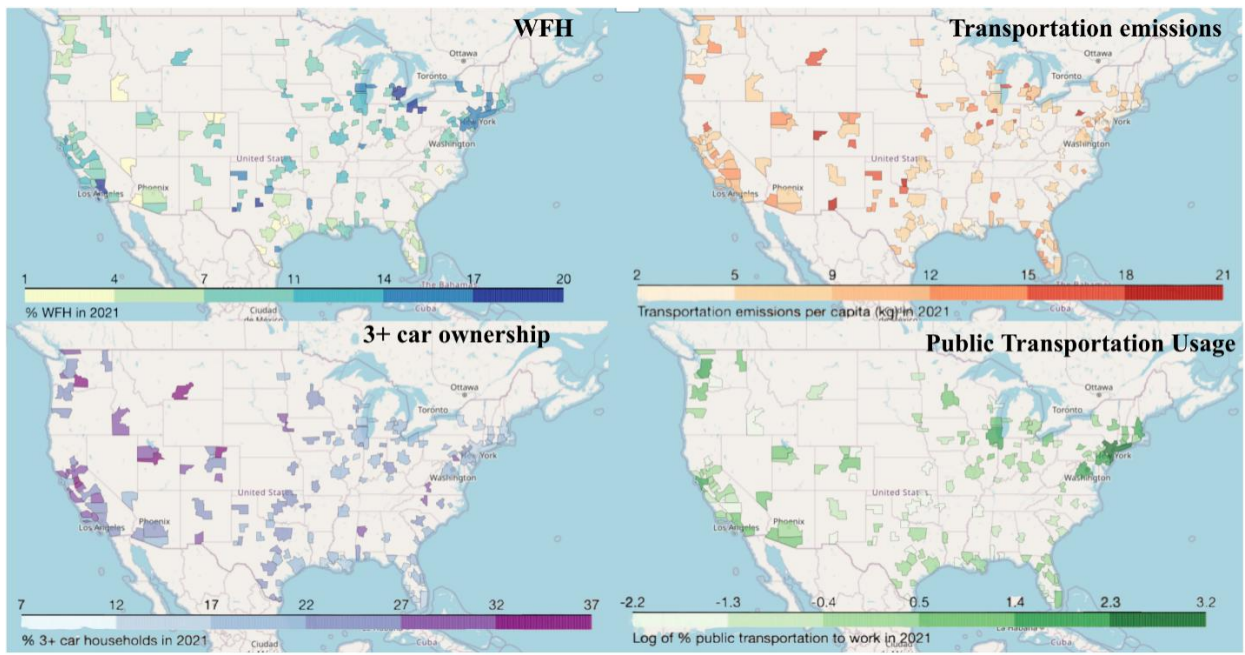

**Figure 1: Distribution of key variables across MSAs in the US in 2021**

The correlation between commute modes across MSAs is significantly high, signifying the sufficiency of including one commute mode in the regression model to avoid multicollinearity issues. For example, in 2019, the percentage of workers taking public transportation to work has a -0.85 correlation with the percentage of workers driving alone and a 0.47 correlation with the percentage of workers walking to work (**Table 5**). We have chosen the percentage of public transportation usage to represent the commute mode. Public transportation was deeply impacted by the pandemic. From 2019 to 2021, the usage of public transportation declined about 40% on average across the 141 MSAs (*6, 7*). Given public transportation's profound change and significant environmental impact, using it as an explanatory variable allows for policy implications that could instigate tangible actions.

*Car Ownership*

*Shen, Wang, Caros, and Zhao*

We use the percentage of households owning at least three cars from the ACS one-year estimates for both 2019 and 2021 to measure car ownership. While the average number of cars per household in the US is around 1.83 in 2022, we want to focus on the tail of the cross-sectional car ownership distribution to capture the dynamics of car-dependent MSAs. It's important to note that about 22% of the households in the US have at least three cars. The empirical results are robust to different definitions of car ownership. This variable helps to provide insights on travel behavior changes in more car dependent MSAs in the US. We find that car ownership is negatively correlated with public transportation usage, positively correlated with driving alone, and not correlated with WFH (**Table 5**). Car ownership not only captures an important added dimension in our multiple regression analysis, but also can initiate important economic and government policy discussions.

**Normality Test**

This potential presence of outliers motivated us to conduct normality tests on both the dependent and explanatory variables. We used the Kolmogorov–Smirnov (K-S) test. The results of the K-S test show the p-values for the percentage of public transportation usage in both 2019 and 2021 to be close to 0, indicating a significant deviation from normality (**Table 4**). Therefore, we use standard log transformation on the percentage of public transportation usage. Other variables all have p-values above 5% in both 2019 and 2021 and, thus, are normally distributed.

**RESULTS**

**Descriptive Statistics**

Table 1 below presents summary statistics for our dependent and explanatory variables in 2019 and 2021. The average public transportation usage across all MSAs was 1.38% in 2019 and declined to 0.84% in 2021, which is approximately a 40 percent decrease. This highlights the unprecedented impact of the pandemic. Concurrently, the percentage of people working from home increased dramatically from a mean of 5.50% in 2019 to 15.23% in 2021. This 177 percent increase underscores the widespread adoption of WFH. The percentage of households with three or more cars remained stable, with a mean of 22.44% in 2019 and 22.41% in 2021.

**Table 1: Descriptive statistics for the modelled variables**

| Variable | Mean | Std. dev | Min | Median | Max | P-value | Unit |
|---|---|---|---|---|---|---|---|
| Transportation emissions per capita, 2019 | 9.92 | 3.79 | 2.5 | 9.24 | 21.81 | 0.34 | Kg $CO_2$ |
| Transportation emissions per capita, 2021 | 9.54 | 3.67 | 2.41 | 8.86 | 20.78 | 0.27 | Kg $CO_2$ |
| Public transportation to work, 2019 | 2.43 | 3.91 | 0 | 1.37 | 33.19 | 0.00 | Pct. |
| Public transportation to work, 2021 | 1.41 | 2.41 | 0 | 0.85 | 24.61 | 0.00 | Pct. |
| Log of public transportation to work, 2019 | 0.32 | 1.05 | -2.3 | 0.32 | 3.5 | 0.69 | Log(Pct.) |
| Log of public transportation to work, 2021 | -0.18 | 1.02 | -2.21 | -0.15 | 3.2 | 0.52 | Log(Pct.) |
| Work From Home (WFH), 2019 | 5.5 | 1.98 | 2.2 | 5.1 | 13.7 | 0.13 | Pct. |
| Work From Home (WFH), 2021 | 15.23 | 6.66 | 3.4 | 14.3 | 36.3 | 0.36 | Pct. |
| 3+ car households, 2019 | 22.44 | 5.13 | 5.8 | 22.2 | 37.4 | 0.39 | Pct. |
| 3+ car households, 2021 | 22.41 | 5.06 | 6.9 | 21.8 | 36.5 | 0.41 | Pct. |

**Regression Analysis**

In order to provide insights into the effects of WFH on transportation emissions in the context of other transportation demand shifts, we use OLS regression.

The WFH variable did not have a statistically significant impact on transportation emissions per capita in both simple and multiple regressions in 2019. However, in 2021, the t-stat for WFH in the simple regression becomes significant at -3.95, indicating a strong relationship where higher WFH adoption is associated with lower emissions. Interpreting the coefficient of simple regression, we observe that a 1% increase in WFH is associated with a 0.18 kg or 1.9% (0.18 kg divided by the mean daily transportation emissions per capita of 2021 at 9.54 kg) reduction in daily average transportation emission per capita in 2021. In the multiple regression in 2021, WFH also becomes a highly significant variable (t-stats -3.13), while controlling for the effect of public transportation and car ownership. Therefore, a 1% increase in WFH was associated with a 0.17 kg or 1.8% reduction of daily average $CO_2$ emissions per capita. The quantification of the impact of WFH on transportation emission is a unique contribution to the literature.

In 2019, increased use of public transportation is significantly associated with lower transportation emissions per capita in both simple and multiple regression results. In simple regression, public transportation usage in 2019 had a t-stat of -4.12. In multiple regression, public transportation has a significant t-stat of -3.81, while controlling for the effects of WFH and car ownership. By reversing the log transformation, we see a 1% increase in the percentage of public transportation usage was associated with a 0.31 kg or 3.2% reduction of daily average transportation emissions per capita. In 2021, public transportation's explanatory power is weakened but still significant (t-stat = -2.38) in simple regression. However, in the multiple regression, public transportation was not statistically significant in 2021, after controlling for WFH and car ownership. This could potentially reflect the result of substantial decline in public transportation use during the pandemic. This indicates that, in relation to other variables, public transportation becomes much less important in influencing transportation emission in 2021.

In 2019, a higher percentage of car ownership, defined as the percentage of households with three or more cars, showed a positively significant association with higher transportation emissions in the simple regression. However, in the multiple regression, the association between car ownership and transportation emissions is not significant in 2019. However, the situation changed in 2021. A higher percentage of car ownership had a significant positive impact on transportation emissions for both simple and multiple regressions. Specifically, our multiple regression analysis reveals that a 1% increase in the percentage of households with three or more cars is associated with a 0.12 kg or 1.3% increase in daily average CO2 emissions per capita. The increased influence of car ownership, in contrast to the declined impact of public transportation, shows the shifting impact of transportation demand on transportation emissions around the pandemic.

*Shen, Wang, Caros, and Zhao***Table 2: Regression and Variance Inflation Factor results**

**Dependent variable: transportation emissions per capita (kg)**

| Simple regression analysis with WFH | 2019 | 2021 |
|---|---|---|
| Constant coefficient | 10.17 | 12.21 |
| *t-stat* | 10.71 | 16.58 |
| WFH coefficient | -0.04 | -0.18 |
| *t-stat* | -0.27 | -3.95 |
| $R^2$ | 0.1% | 10.1% |
| **Simple regression analysis with public transportation** | **2019** | **2021** |
| Constant coefficient | 10.31 | 9.40 |
| *t-stat* | 32.55 | 30.53 |
| Log of public transportation coefficient | -1.19 | -0.76 |
| *t-stat* | -4.11 | -2.56 |
| $R^2$ | 10.8% | 4.5% |
| **Simple regression analysis with 3+ car households** | **2019** | **2021** |
| Constant coefficient | 7.28 | 6.26 |
| *t-stat* | 5.12 | 4.52 |
| 3+ car household coefficient | 0.12 | 0.15 |
| *t-stat* | 1.91 | 2.43 |
| $R^2$ | 2.6% | 4.1% |
| **Multiple regression** | **2019** | **2021** |
| Constant coefficient | 8.25 | 9.34 |
| *t-stat* | 5.09 | 5.94 |
| Log of public transportation coefficient | -1.18 | 0.02 |
| *t-stat* | -3.81 | 0.06 |
| WFH coefficient | 0.13 | -0.17 |
| *t-stat* | 0.79 | -3.13 |
| 3+ car household coefficient | 0.06 | 0.12 |
| *t-stat* | 1.00 | 2.02 |
| $R^2$ | 11.9% | 12.9% |
| **Variance Inflation Factor (VIF)** | **2019** | **2021** |
| Constant | 28.70 | 29.03 |
| Log of public transportation | 1.15 | 1.58 |
| WFH | 1.08 | 1.46 |
| 3+ car households | 1.06 | 1.10 |

The R² values for 2019 and 2021 are 11.92% and 12.86%, respectively, indicating a modest level of explanatory power. We checked the multicollinearity by calculating the Variance Inflation Factor (VIF) for the 2019 and 2021 models. No issues have been identified. In addition, we observe that the correlations across the three chosen explanatory variables are reasonably low in 2019 and 2021, as shown in Table 2.

**Blinder-Oaxaca Decomposition**

We have observed in Table 3 that transportation emission per capita has declined by 0.38 kg from 2019 to 2021. We use Blinder-Oaxaca decomposition to get a detailed decomposition of the endowment effect, the coefficient effect, and the interaction effect. The endowment effect measures the decline of transportation emissions that is due to the difference in the levels of three explanatory variables (WFH, Public Transportation and Car Ownership) from 2019 to 2021. The coefficient effect measures the impact on the decline of transportation emissions due to differences in the estimated coefficients of the three explanatory variables in the 2019 vs. 2021 multiple regressions. The coefficient effect can serve as a proxy for unobserved structural changes around the pandemic. Lastly, the interaction effect measures the contribution to the decline in transportation emission from 2019 to 2021 due to the combined influence of differences in both endowment and coefficient effects. The interaction effect arises because endowment effects and coefficient effects are often not independent of one another.

**Table 3: Blinder-Oaxaca decomposition results (2021 vs. 2019)**

| **Explanatory variables** | **Endowment** | **Coefficient** | **Interaction** | *Row sum* |
|---|---|---|---|---|
| Constant | 0.00 | 1.08 | 0.00 | 1.08 |
| Log of public transportation | 0.60 | 0.39 | -0.61 | 0.38 |
| WFH | 1.23 | -1.61 | -2.84 | -3.22 |
| 3+ car households | 0.00 | 1.38 | 0.00 | 1.38 |
| *Column sum* | 1.82 | 1.25 | -3.45 | -0.38 |

Note: Transportation emissions per capita (kg) were used as the dependent variable.

From the right side column (Row Sum) of Table 3, we can see that among the explanatory variables, WFH had the most significant impact on reducing transportation emissions, with a total impact of -3.22. WFH has a significantly negative coefficient effect at -1.61. The regression coefficient for WFH in 2021 turned significantly negative, indicating that more WFH leads to less transportation emissions per capita. Therefore, the increase in WFH from 2019 to 2021 leads to a strong coefficient effect of reduction in transportation emissions per capita.

Counterbalancing the reductive effects from change in WFH are both public transportation and car ownership. Public transportation has a positive contribution to increase the transportation emissions from 2019 to 2021. Importantly, public transportation has an endowment effect of 0.60. Public transportation declined by 40% on average from 2019 to 2021. This decline is then multiplied by the reductive effect of public transportation on transportation emissions, as seen in the coefficient of multiple regression (-1.18) in 2019 (Table 2). Therefore, the decomposition results show that the decline in public transportation usage has led to an increase in transportation emissions per capita from 2019 to 2021.

Car ownership has a significantly positive contribution (+1.38) to increase the transportation emissions, we can see that this is entirely driven by its coefficient effect. Therefore, even though three + car ownership didn't change significantly from 2019 to 2021, it becomes a more significantly positive contributor to the increase in transportation emissions from 2019 to 2021. The unobserved structural effect, highlighted by coefficient effect, shows the importance of car ownership to increase transportation emissions from 2019 to 2021.

We note that most components of the decomposition have decomposition shares greater than 100%, indicating the dramatic changes during the pandemic. From the bottom row (Column Sum) of Table 3, we observe that both the endowment and coefficient effects have a positive impact on transportation emissions per capita while the interaction effect has a large reductive impact on transportation emissions at -3.45. The substantial changes in the values of the explanatory variables between the two periods result in a large endowment effect. The coefficients for WFH and public transportation in the multiple regressions in 2019 and 2021 show marked differences, leading to large coefficient effects. The interdependence of the endowment and coefficient effects have resulted in the large interaction effect.

In summary, the decomposition analysis underscores the significant role of increased WFH adoption in reducing transportation emissions. At the same time, the reductive effect of public transportation usage on transportation emissions has decreased due to the decline of public transportation during the pandemic. The impact of car ownership on increased emissions has grown. Therefore, a careful evaluation of WFH's reductive impact on transportation emissions needs to account for the counterbalancing effect from public transportation and car ownership.

## DISCUSSION

### The Impact of WFH on Public Transportation and Car Ownership

There are multiple pathways for WFH to impact transportation emissions as we have seen through the simple and multiple regressions. Next, we aim to identify the relationship between the changes in WFH and the changes in public transportation and car ownership, respectively. It is important to note that Zheng et al. (22) have found causality between the rise of WFH and the decline of public transportation ridership; our findings are consistent.

We find that the increase of WFH is significantly associated with the decline in public transportation with a t-stat of -9.56 (see Table 4). We see that a 1% increase in WFH corresponds to a 0.21% reduction in public transportation usage to work. At the same time, the increase of WFH had a marginal impact on car ownership with an insignificant t-stat.

**TABLE 4. Impact of WFH on public transportation and car ownership**

| **Dependent variable: Change in public transportation, 2019 to 2021** | | |
|---|---|---|
| | Coefficient | t-stat |
| Constant | 1.02 | 4.16 |
| Change in WFH from 2019 to 2021 | -0.21 | -9.56 |
| $R^2$ | 39.7% | |
| **Dependent variable: Change in 3+ car households from 2019 to 2021** | | |
| | Coefficient | t-stat |
| Constant | 0.39 | 1.32 |
| Change in WFH from 2019 to 2021 | -0.04 | -1.61 |
| $R^2$ | 1.8% | |

**Excluded Variables**

In our multiple regression analysis, we considered but ultimately excluded five notable variables: population, remote-ready industry, mean travel time to work, household income, and average household car ownership.

Population is excluded because its negative, nonlinear relationship with transportation emissions per capita is well-documented and mostly driven by larger cities with more developed public transportation systems, which tend to have lower emissions per capita. Also, population has a high correlation with public transportation usage (0.73 in 2019 and 0.77 in 2021, **Table 5**), leading to potential multicollinearity.

The remote-ready industry variable, encompassing jobs inherently suitable for remote work, is created by aggregating the percentage of jobs in information technology, finance, real estate, insurance, and professional or managerial positions. This variable has a high correlation with % WFH (0.50 in 2019 and 0.82 in 2021, **Table 5**). We chose to exclude the remote ready industry variable due to potential multicollinearity and because it is a less direct measure of WFH.

Mean travel time to work was excluded due to its composite nature, which includes all commute modes and is influenced by factors such as population density and transportation infrastructure quality. This variable correlates positively with public transportation usage and negatively with driving alone, making it a complex approximation for commute mode choices.

Household income was excluded due to its multifaceted impact on transportation demand. Higher household income can lead to increased mobility and more flexible transportation choices, complicating its relationship with transportation emissions.

Lastly, average household car ownership is estimated by taking the weighted average of the percentage of households with zero, one, two, and more than three cars. We find that this variable has a significantly positive relationship with the percentage of households owning more than three cars (0.93 in 2019 and 0.92 in 2021, as shown in Table 5). The results are found to be consistent whether we use the percentage of households owning at least three cars or other car ownership definitions (percentage of households owning at least one car, at least two cars, or average car ownership per household) as independent variables in the model. See more in the appendix.

In conclusion, there are clear trade-offs to these exclusions. Instead of pursuing a general understanding of the drivers behind transportation emissions, we aim to provide a focused event analysis of the impact of WFH on transportation emissions during the pandemic. Our choices aim to balance the inclusion of the relevant variables with the considerations of statistical power, multicollinearity, and focus on the effect of the socioeconomic variables with the most changes during the pandemic.

**Table 5: Correlation between potential explanatory variables in 2019 and 2021**

| Explanatory variable | Pop. | Drive alone to work | Carpool to work | Public transit to work | Walk to work | Other means to work | WFH | Remote-ready jobs | 3+ car HHs | Avg. car per HH |
|---|---|---|---|---|---|---|---|---|---|---|
| Population | 1.00 | -0.56 | 0.00 | 0.77 | 0.16 | 0.19 | 0.42 | 0.52 | -0.23 | -0.39 |
| Drive alone to work | -0.54 | 1.00 | -0.32 | -0.77 | -0.64 | -0.59 | -0.45 | -0.36 | 0.15 | 0.35 |
| Carpool to work | -0.14 | -0.21 | 1.00 | -0.11 | -0.18 | 0.09 | -0.12 | -0.09 | 0.19 | 0.20 |
| Public transit to work | 0.73 | -0.85 | -0.16 | 1.00 | 0.44 | 0.25 | 0.41 | 0.40 | -0.29 | -0.54 |
| Walk to work | 0.14 | -0.65 | -0.11 | 0.47 | 1.00 | 0.31 | 0.38 | 0.18 | -0.12 | -0.24 |
| Other means to work | 0.05 | -0.44 | -0.05 | 0.20 | 0.25 | 1.00 | 0.39 | 0.32 | -0.10 | -0.14 |
| WFH | 0.11 | -0.23 | -0.07 | 0.15 | 0.07 | 0.46 | 1.00 | 0.82 | -0.12 | -0.12 |
| **Remote-ready jobs** | 0.50 | -0.33 | -0.12 | 0.44 | 0.11 | 0.02 | 0.50 | 1.00 | -0.23 | -0.21 |
| **3+ car HHs** | -0.21 | 0.08 | 0.39 | -0.23 | -0.15 | -0.08 | -0.01 | -0.09 | 1.00 | 0.92 |
| **Avg. car per HH** | -0.36 | 0.29 | 0.37 | -0.46 | -0.27 | -0.09 | 0.09 | -0.05 | 0.93 | 1.00 |

## CONCLUSIONS

Our study reveals a multifaceted impact of WFH on transportation emissions over the COVID-19 pandemic period. WFH showed limited influence on transportation emissions before the pandemic. The widespread adoption of WFH since the pandemic resulted in a substantial reductive impact in transportation emissions, highlighting its potential to contribute positively to environmental goals. We contribute to the literature by quantitatively estimating the impact of WFH to transportation emissions reduction. After controlling for public transportation usage and car ownership, we find that a 1% increase in WFH corresponds to a 0.17 kg or 1.8% reduction of daily average transportation emissions per capita.

Despite the pronounced decrease in transportation emissions due to WFH, our findings also underscore critical trade-offs. Our findings show that public transportation, recognized for its impact on reducing transportation emissions, became less effective in the context of an unprecedented drop in ridership and increased WFH.

Furthermore, our study identifies an evolving trend in car ownership. Although car ownership did not significantly change during the pandemic, we find it to have a more significant impact on increasing transportation emissions in 2021, after controlling for the effect of WFH and public transportation. The concurrent increase in car ownership's impact on emissions, alongside the decline of public transportation usage and the rise of WFH adoption, reflects the multifaceted impact of pandemic-era travel behavior changes on transportation emissions.

### Policy Implications

Our research highlights the importance of a nuanced and balanced approach in shaping WFH policies. From existing literature, it's important to note the side effects and derived transportation behavior induced by WFH. Third places, additional leisure travel, donut effect from urban planning, and reorienting public transportation from commuting to leisure travel are some of the important topics of research. We add to the debate on WFH by advocating for WFH to reduce transportation emissions while mitigating unintended consequences. Policymakers can develop strategies that maximize emissions reductions while mitigating unintended consequences. Effective policy interventions should prioritize promoting public transportation alongside WFH initiatives. Concurrently, encouraging the adoption of environmentally friendly commute modes, such as cycling and walking, can complement WFH by further reducing reliance on private vehicles. Our study advocates for a data-driven and holistic approach to shape future WFH policies. By addressing the complex interactions between WFH, transportation behavior, and emissions outcomes, policymakers can harness WFH as a powerful tool for advancing environmental sustainability goals.

**Future Research**

Our research design focuses on studying specific dimensions of transportation demand within the context of a particular time period: the COVID-19 pandemic. This focused approach resembles an event study, offering valuable insights into the effects of disruptive events like COVID-19. Future research extensions could leverage the real-time, granular aspects of the Carbon Monitor Cities dataset to provide deeper insights into temporal and spatial dynamics of transportation emissions. Transportation emissions could also be influenced by a host of additional variables such as commute distance, travel frequency, vehicle type, joint-purpose trip and others. A further research direction could be a comprehensive study investigating the combined effects of commute distance, travel frequency, and vehicle type on transportation emissions. In addition, The temporal variation could be of future research focus when the emission data becomes available for 2022 and beyond. Given the global coverage of emissions data from CMCs coupled with socio-economic data like ACS around the world, we can extend the study to a global scale.

Moreover, employing advanced analytical methods such as Structural Equation Modeling (SEM) could elucidate the complex relationships between explanatory variables and transportation outcomes. Complementing our targeted approach, machine learning techniques, such as Lasso or Ridge regression could also aid in variable/feature selection, offering empirical insights into the complex relationships involved.

**Appendix**

By substituting the percentage of households with three or more cars variable in the multiple regression with the average car ownership per household, one or more and two or more car ownership households, we find the key findings stay consistent. Our key quantification of the impact of WFH on transportation emission per capita is -0.17 kg in 2021 independent of the definition of car ownership. The general increasing pattern of the t-statistics of car ownership from 2019 to 2021 is also the same with different definitions of car ownership.

| Dependent variable: transportation emissions per capita (kg) | 3+ car Households | | 2+ car Households | | 1+ car Households | | Avg Cars per Household | |
|---|---|---|---|---|---|---|---|---|
| **Multiple regression** | **2019** | **2021** | **2019** | **2021** | **2019** | **2021** | **2019** | **2021** |
| const | 8.25 | 9.34 | 9.67 | 7.43 | 10.18 | 1.27 | 7.82 | 4.16 |
|  | 5.09 | 5.94 | 2.91 | 2.27 | 0.85 | 0.11 | 1.71 | 0.91 |
| Log of public transportation | -1.18 | 0.02 | -1.26 | 0.05 | -1.27 | 0.05 | -1.21 | 0.12 |
|  | -3.81 | 0.06 | -3.84 | 0.13 | -3.34 | 0.12 | -3.62 | 0.31 |
| WFH | 0.13 | -0.17 | 0.14 | -0.17 | 0.14 | -0.17 | 0.12 | -0.17 |
|  | 0.79 | -3.13 | 0.83 | -3.18 | 0.79 | -3.11 | 0.74 | -3.23 |
| Car Ownership (different definitions) | 0.06 | 0.12 | 0 | 0.08 | 0 | 0.12 | 1.04 | 4.57 |
|  | 1 | 2.02 | -0.03 | 1.43 | -0.05 | 0.9 | 0.39 | 1.73 |
| R2 | 0.119 | 0.129 | 0.112 | 0.116 | 0.113 | 0.108 | 0.114 | 0.122 |

## ACKNOWLEDGMENTS

The authors are grateful for feedback and support from the students and faculty of the MIT JTL Urban Mobility Lab.

## AUTHOR CONTRIBUTIONS

*Shen, Wang, Caros, and Zhao*

22